\documentclass[twocolumn,nofootinbib,amsmath,amssymb,superscriptaddress,
preprintnumbers,a4paper]{revtex4}

\usepackage{graphicx}% Include figure files
\usepackage{dcolumn}% Align table columns on decimal point
\usepackage{bm}

\newcommand{\be}{\begin{equation}}
\newcommand{\ee}{\end{equation}}
\def\ltorder{\mathrel{\raise.3ex\hbox{$<$}\mkern-14mu
 \lower0.6ex\hbox{$\sim$}}}

\begin{document}

% Setup title page
\preprint{
\vbox{
\hbox{SHEP-0706} 
\hbox{Edinburgh 2007/4}
}}
%\preprint{Edinburgh 2007/4}

\title{$K_{l3}$ form factor with $N_f=2+1$ dynamical domain wall
  fermions: A progress report}% Force line breaks with \\

\author{D.~J.~Antonio}
\affiliation{School of Physics, University of Edinburgh,
  Edinburgh EH9 3JZ, UK}
\author{P.~A.~Boyle}
\affiliation{School of Physics, University of Edinburgh,
  Edinburgh EH9 3JZ, UK}
\author{C.~Dawson}
\affiliation{RIKEN-BNL Research Center, Brookhaven National
  Laboratory,  Upton, NY 11973, USA}
\author{T.~Izubuchi}
\affiliation{RIKEN-BNL Research Center, Brookhaven National
  Laboratory,  Upton, NY 11973, USA}
\affiliation{Institute for Theoretical Physics, Kanazawa University,
  Kanazawa, Ishikawa 920-1192, Japan}
\author{A.~J\"uttner}
\affiliation{School of Physics and Astronomy, University of
  Southampton,  Southampton, SO17 1BJ, UK}
\author{C.~Sachrajda}
\affiliation{School of Physics and Astronomy, University of
  Southampton,  Southampton, SO17 1BJ, UK}
\author{S.~Sasaki}
\affiliation{RIKEN-BNL Research Center, Brookhaven National
  Laboratory,  Upton, NY 11973, USA}
\affiliation{Department of Physics, University of Tokyo, Tokyo
  113-0033,  Japan}
\author{A.~Soni}
\affiliation{Physics Department, Brookhaven National Laboratory,
  Upton, NY 11973, USA}
\author{R.~J.~Tweedie}
\affiliation{School of Physics, University of Edinburgh,
  Edinburgh EH9 3JZ, UK}
\author{J.~M.~Zanotti}
 \thanks{Talk presented by J.~Zanotti at CKM2006, Nagoya, Japan}
\affiliation{School of Physics, University of Edinburgh,
  Edinburgh EH9 3JZ, UK}
\collaboration{UKQCD/RBC Collaboration} \noaffiliation

\begin{abstract}
  We present the latest results from the UKQCD/RBC
  collaborations for the $K_{l3}$ form factor with $2+1$ flavours of
  dynamical domain wall quarks. Simulations are performed on
  $16^3\times 32\times 16$ and $24^3\times 64\times 16$ lattices with
  three values of the light quark mass, allowing for an extrapolation
  to the chiral limit.
  After interpolating to zero momentum transfer, we obtain the
  preliminary result $f_+^{K\pi}(0)=0.9609(51)$ (or $\Delta f =
  -0.0161(51)$), which is in agreement with the result of Leutwyler \&
  Roos.
\end{abstract}

\maketitle

\section{INTRODUCTION}
\vspace*{-2mm}

$K\rightarrow \pi l\nu\ (K_{l3})$ decays provide an excellent avenue
for an accurate determination of the Cabibbo-Kobayashi-Maskawa (CKM)
\cite{Cabibbo:1963yz} quark mixing matrix element, $|V_{us}|$.
This is done by observing that the decay amplitude is proportional to
$|V_{us}|^2 |f_+(q^2)|^2$, where $f_+(q^2)$ is the form factor
defined from the $K\to \pi$ matrix element of the weak vector current,
$V_\mu=\bar{s}\gamma_\mu u$
\be
\langle \pi(p^\prime) \big | V_\mu \big | K(p)\rangle = (p_\mu +
p_\mu^\prime) f_+(q^2) + (p_\mu - p_\mu^\prime) 
f_-(q^2)\,,
\label{eq:ME}
\ee
where $q^2=(p-p^\prime)^2$.

In chiral perturbation theory (ChPT), $f_+(0)$ is expanded in terms of
the light pseudoscalar meson masses, $m_\pi,\,m_K,\,m_\eta$
\be
f_+(0) = 1 + f_2 + f_4 + \ldots,\quad (f_n={\cal
  O}(m^n_{\pi,\,K,\,\eta}))\ .
\ee
Conservation of the vector current ensures that $f_+(0) = 1$ in the
$SU(3)$ flavour limit.
Additionally, as a result of the Ademollo-Gatto Theorem
\cite{Ademollo:1964sr}, which states that $f_2$ receives no
contribution from local operators appearing in the effective theory,
$f_2$ is determined unambiguously in terms of $m_\pi$, $m_K$ and
$f_\pi$, and takes the value $f_2=-0.023$ at the physical values of
the meson masses \cite{Leutwyler:1984je}.

Our task is now reduced to one of finding
\be
\Delta f = f_+(0) - (1 + f_2) \ .
\label{eq:deltaf}
\ee
In order to obtain a result for $f_+(0)$ which is accurate to
$\sim$1\%, it is sufficient to have a 20-30\% error on $\Delta f$.
Until recently, the standard estimate of $\Delta f=-0.016(8)$ was due
to Leutwyler \& Roos \cite{Leutwyler:1984je}, however a more recent
ChPT analysis favours a positive value, $\Delta f=0.007(12)$
\cite{Cirigliano:2005xn}.
A calculation of $\Delta f$ on the lattice is therefore essential.

The last few years have seen an improvement in the accuracy of lattice
calculations of this quantity
\cite{Becirevic:2004ya,Okamoto:2004df,Tsutsui:2005cj,Dawson:2005zv,Dawson:2006qc},
with the results favouring a negative value for $\Delta f$ in
agreement with Leutwyler \& Roos.

The UKQCD and RBC collaborations have embarked on a program to perform
the first unitary (i.e. $N_f=2+1$ flavour) lattice calculation of the
$K\to\pi$ form factor using dynamical domain wall fermions at light
quark masses and on large volumes.
Preliminary results from this study were introduced in
Ref.~\cite{Antonio:2006ev}.
The work presented here represents an improvement on the earlier
analysis through a more in depth analysis of the systematic errors
involved in the extraction of $f_+(0)$ from lattice correlation
functions.

\vspace*{-2mm}
\section{Lattice Techniques}
\vspace*{-2mm}
\subsection{Parameters}
\vspace*{-2mm}

We simulate with $N_f=2+1$ dynamical flavours generated with the
Iwasaki gauge action \cite{Iwasaki:1985we} at $\beta=2.13$, which
corresponds to an inverse lattice spacing $a^{-1}\approx
1.62(4)\,\text{GeV}$ ~\cite{Allton:2007hx}, and the domain wall
fermion action \cite{Kaplan:1992bt} with domain wall height $M_5=1.8$
and fifth dimension length $L_s=16$.
This results in a residual mass of $am_\text{res}\approx 0.00308(4)$
\cite{Allton:2007hx}.
The simulated strange quark mass, $am_s=0.04$, is very close to its
physical value \cite{Allton:2007hx}, and we choose three values for
the light quark masses, $am_{ud}=0.03,\,0.02,\,0.01$, which correspond
to pion masses $m_\pi\approx 630$, 520, 390 MeV \cite{Allton:2007hx}.
The calculations are performed on two volumes, $16^3\times 32$ and
$24^3\times 64$, at each quark mass.
See \cite{Allton:2007hx} for further simulation details.

\vspace*{-2mm}
\subsection{Extracting form factors}
\label{sec:formfac}
\vspace*{-2mm}

\begin{figure}[tb]
\vspace*{-7mm}
\includegraphics[width=6cm,angle=-90]{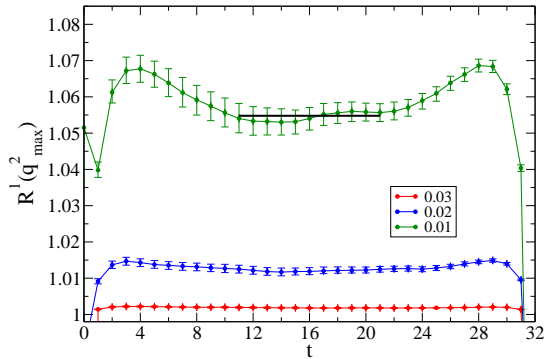}
\vspace*{-6mm}
\caption{Ratio for $f_0(q_\text{max}^2)$, $R(t',t)$, as defined in
  Eq.~({\protect\ref{eq:ratio1}}), for three simulated light masses
  $am_{ud}=0.03,\,0.02,\,0.01$ on a $24^3\times 64$ volume. Further
  simulation parameters can be found in
  ~{\protect\cite{Allton:2007hx}}.}
\label{fig:ratio1}
\vspace*{-3mm}
\end{figure}

The techniques we use to calculate $f_+(0)$ are similar to those
originally proposed in \cite{Becirevic:2004ya} and have been outlined
in detail in Ref.~\cite{Dawson:2006qc}.
We restrict ourselves here to highlighting the main points.

We start by rewriting the vector form factors given in
Eq.~(\ref{eq:ME}) to define the scalar form factor
\be
f_0(q^2) = f_+(q^2) + \frac{q^2}{m_K^2 - m_\pi^2}f_-(q^2)\ ,
\ee
which can be obtained on the lattice at $q^2_\text{max}=(m_K -
m_\pi)^2$ with high precision from the following ratio
\cite{Hashimoto:1999yp}
\begin{eqnarray}
R(t',t) &=& \frac
{C^{K\pi}_4(t',t;\vec{0},\vec{0})C^{\pi K}_4(t',t;\vec{0},\vec{0})}
{C^{KK}_4(t',t;\vec{0},\vec{0})C^{\pi\pi}_4(t',t;\vec{0},\vec{0})}
\nonumber\\
&\xrightarrow[t,(t'-t) \to \infty]{}& \frac{(m_K +
  m_\pi)^2}{4m_K m_\pi} \big| f_0(q^2_\text{max})\big|^2\ ,
\label{eq:ratio1}
\end{eqnarray}
where the three-point function is defined as
\begin{eqnarray}
\lefteqn{C_{\mu}^{PQ}(t', t,\vec{p}\,',\vec{p})  
   = \sum_{\vec{x}, \vec{y}}
   e^{-i\vec{p}\,'(\vec{y}-\vec{x})} e^{-i\vec{p} \vec{x}} \times}
   \\
&&\big < 0 \big |  {\cal O}_Q(t',\vec{y})\big |
   Q \big > \big < Q \big |
   V_\mu(t,\vec{x}) \big | P \big > \big < P \big | 
   {\cal O}^\dagger_P(0) \big | 0 \big > \ ,
\nonumber
\end{eqnarray}
with $P,Q=\pi$ or $K$ and ${\cal O}_{\pi(K)}$ is a local interpolating
operator for a pion(kaon).
We note that $R(t',t)=1$ in the $SU(3)_\text{flavour}$ symmetric
limit, hence any deviations from unity are purely due to
$SU(3)_\text{flavour}$ symmetry breaking effects.

In Fig.~\ref{fig:ratio1} we display our
results for $R(t',t)$ for each of the simulated quark masses as
obtained on the $24^3\times 64$ lattices.

It is immediately obvious that $f_0(q^2_\text{max})$ can be measured
with a very high level of statistical accuracy.
We also note that the ratio becomes larger the further we move away
from the $SU(3)_\text{flavour}$ limit.
Since there is no spatial momentum involved in this ratio, the results
obtained on the two different volumes should agree, and any difference
can only be due to finite size effects.
We find that within statistical errors, finite size effects on
$f_0(q^2_\text{max})$ are negligible~\cite{Antonio:2006ev}.

In order to to extract the form factor at $q^2=0$, we need to obtain
results at finite $q^2<0$ and by studying the $q^2$ dependence we are
then in a position to interpolate to $q^2=0$.
We calculate the form factor $f_0(q^2<0)$ using the ratio method
outlined in Ref.~\cite{Becirevic:2004ya,Dawson:2006qc} which has the
advantage that no knowledge of $Z_V$ is required.
We simulate with two unique choices of the three-momentum transfer,
$\vec{q}=2\pi/L(1,0,0)$ and $\vec{q}=2\pi/L(1,1,0)$, where $L$ is the
spatial extent of our lattice.
We take advantage of rotational symmetry to improve the signal,i.e.
we use all 6 and 12 permutations of (1,0,0) and (1,1,0), respectively.
The ratio method also allows $f_0(q^2)$ to be extracted from the
$\pi\to K$ matrix element which, due to the difference between the
$\pi$ and $K$ masses, doubles the number of $q^2$ values.
Finally, since (spatial) momentum on the lattice is proportional to
the inverse (spatial) lattice size, our simulations on two different
volumes provides access to additional $q^2$.

\vspace*{-3mm}
\section{Results}
\vspace*{-2mm}
\subsection{Interpolation to $q^2=0$}
\vspace*{-2mm}

\begin{figure}[tb]
\vspace*{-1mm}
\includegraphics[width=7.5cm]{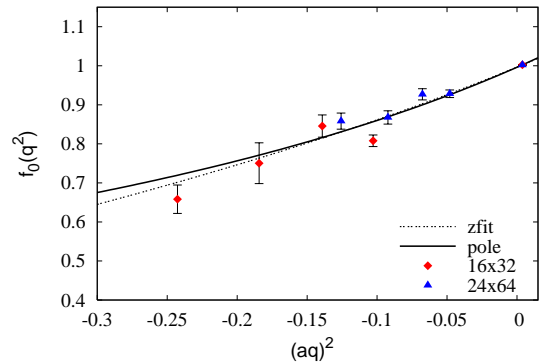}
\vspace*{-4mm}
\caption{Scalar form factor $f_0(q^2)$ for bare quark mass
  $am_{ud}=0.02$. Results are obtained on two
  volumes $V=16^3\times 32$ (red diamonds) and $V=24^3\times 64$ (blue
  triangles). The solid line is the result of a fit using
  a monopole ansatz (Eq.~({\protect\ref{eq:monopole}})) while the dotted line is the result of a fit using a zfit ansatz (Eq.~({\protect\ref{eq:zfit}})).}
\label{fig:fzero}
\vspace*{-2mm}
\end{figure}

We present our results obtained on each volume for $f_0(q^2)$ in
Fig.~\ref{fig:fzero} for quark mass $am_{ud}=0.02$.
In the intermediate $q^2$-range we see good agreement
between the results obtained on the two different volumes, indicating
that finite size effects are negligible, at least for the quark masses
considered here.
This means that we now have results over a large range of $q^2$ to fit
to.

It is not immediately clear what is the best method for fitting $f_0(q^2)$,
however previous work \cite{Dawson:2006qc,Becirevic:2004ya} has found that
lattice results are described very well by a monopole form
\be
f_0(q^2) = \frac{f_0(0)}{(1-q^2/M^2)}\ ,
\label{eq:monopole}
\ee
where $M$ is the monopole mass.
From Fig.~\ref{fig:fzero} we see that our data is also described well by
Eq.~\ref{eq:monopole} (solid line).
From this fit, we are able to calculate $f_0(0)$ for our three simulated quark
masses.

\vspace*{-2mm}
\subsection{Chiral Extrapolation}
\vspace*{-2mm}

Now that we have obtained results for $f_+(0)=f_0(0)$ at three
different quark masses, we are in a position to attempt an
extrapolation to the physical pion mass.
Inserting these results into the expression given in
Eq.~(\ref{eq:deltaf}), together with $f_2$ calculated at the simulated
quark masses using the ChPT formula
\cite{Leutwyler:1984je,Becirevic:2005py}, we are now left with the 
task of chirally extrapolating $\Delta f$.

The Ademollo-Gatto Theorem implies that to leading order $\Delta f \propto (m_s - m_{ud})^2$, hence we attempt a chiral extrapolation using
\be
\Delta f = a + B(m_s - m_{ud})^2\ .
\label{eq:chiral1}
\ee
Note that in the $SU(3)_\text{flavour}$ limit, $\Delta f=0$, so we
expect that a fit to our data should produce $a\approx 0$, and indeed
we find this to be the case.
In the chiral limit we find $\Delta f =
-0.0146(28)$.

\begin{figure}[tb]
\vspace*{-1mm}
\hspace*{-6mm}
\includegraphics[width=7.5cm]{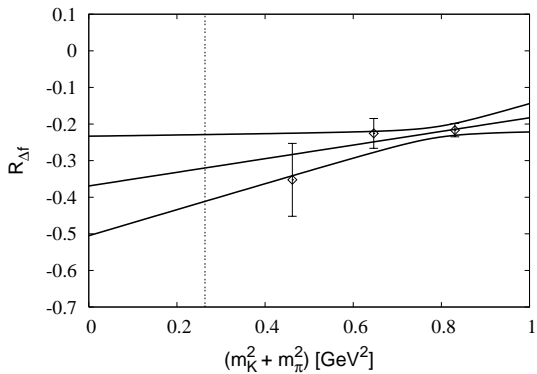}
\vspace*{-3mm}
\caption{Chiral extrapolation of $\Delta f$ using   
  Eq.~({\protect\ref{eq:chiral2}}).
}
\label{fig:chiral}
\vspace*{-2mm}
\end{figure}

In order to attempt to take into account higher terms in the chiral expansion of
$\Delta f$, it has been noted that it is convenient to consider an
extrapolation of the ratio \cite{Becirevic:2004ya,Dawson:2006qc}
\be
R_{\Delta f} = \frac{\Delta f}{(m_K^2 - m_\pi^2)^2} = a + b(m_K^2 +
m_\pi^2)\ .
\label{eq:chiral2}
\ee
We show the extrapolation using this form in Fig~\ref{fig:chiral} from
which we extract a result at the physical meson masses (vertical
dotted line), $\Delta f =-0.0161(46)$.

\vspace*{-2mm}
\subsection{Estimating Systematic Errors}
\vspace*{-2mm}

As we have seen in the previous section, with only three quark masses it is
difficult to distinguish between different chiral extrapolation ans\"atze.
Hence, we quote the result obtained using Eq.~(\ref{eq:chiral2}), and use
the difference between the results obtained from the two
extrapolations (0.0015) as an estimate of the systematic error due to
the chiral extrapolation.

Another possible source of systematic error comes from our choice of a
monopole fit for the $q^2$ dependence of $f_0(q^2)$.
In order to estimate the systematic error due to this choice, we fit
$f_0(q^2)$ at each quark mass with a linear form
\be
f_0(q^2)=f_0(0) + a_1 q^2\,,
\label{eq:linearfit}
\ee
a quadratic form
\be
f_0(q^2)=f_0(0) + a_1 q^2 + a_2 q^4\,,
\label{eq:quadfit}
\ee
and a parameterisation proposed in Ref.~\cite{Hill:2006bq}
\be
f_0(t)=\frac{1}{\phi(t,t_0,Q^2)}\sum_{k=0}^\infty a_k(t_0,Q^2) z(t,t_0)^k\,,
\label{eq:zfit}
\ee
where $|z|<1$ is a mapping of $t=q^2$ onto the unit circle in the
complex plane
$$
t\to z(t,t_0)\equiv\frac{\sqrt{t_+-t}-
  \sqrt{t_+-t_0}}{\sqrt{t_+-t}+\sqrt{t_+-t_0}}\,,
$$
and $t_\pm\equiv(m_K\pm m_\pi)^2$. 
We use $t_0=t_+(1-\sqrt{1-t_-/t_+})$ as the point that maps onto $z=0$, and\\
\begin{eqnarray*}
\phi(t,t_0,Q^2)&=&\sqrt{\frac{3t_+t_-}{32\pi}} \frac{z(t,0)}{-t}
\frac{z(t,-Q^2)}{-Q^2-t} \left(\frac{z(t,t_0)}{t_0-t} \right)^{-1/2}\\
&\times& \left(\frac{z(t,t_-)}{t_- -t} \right)^{-1/4}
\frac{\sqrt{t_+-t}}{(t_+-t_0)^{1/4}}\, ,
\end{eqnarray*}
with $Q^2=2/a^2$ and $a^{-1}=1.62$~GeV is the lattice spacing.
We note that employing different values for $t_0$ and $Q^2$ simply
affects the fitted values of the coefficients, $a_k$, and not the
result for $f_0(0)$.

As an example, the dotted line in Fig.~\ref{fig:fzero} shows the
result of a fit to $f_0(q^2)$ for quark mass $am_{ud}=0.02$ using the
parameterisation in Eq.~(\ref{eq:zfit}) truncated to include terms up
to $k=2$.

We find that all four parameterisations describe our data reasonably
well, except the linear ansatz which leads to a large $\chi^2/dof$ on
the $am=0.01$ dataset.
We take the difference between the result from the pole fit and the z
fit as the estimate of the systematic error due to the $q^2$
dependence.

\begin{figure}[tb]
\vspace*{-1mm}
\includegraphics[width=7.5cm]{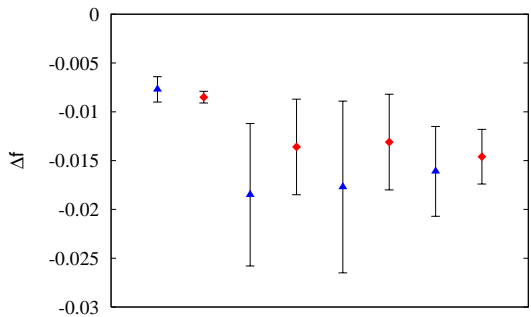}
\vspace*{-9mm}
\caption{Comparison of $\Delta f$ at the physical meson masses
  from all four parameterisations.
  Blue triangles indicate the results using Eq.~(\ref{eq:chiral1}) for
  the chiral extrapolation, while the red diamonds use
  Eq.~(\ref{eq:chiral2}).  The four sets of points are, moving from
  left to right, linear, quadratic, z and pole fits, respectively.}
\label{fig:comparison}
\vspace*{-3mm}
\end{figure}

In Fig.~\ref{fig:comparison}, we provide a comparison of the value of
$\Delta f$ at the physical meson masses as obtained from all four
parameterisations.
The blue triangles indicate the results using Eq.~(\ref{eq:chiral1})
for the chiral extrapolation, while the red diamonds use
Eq.~(\ref{eq:chiral2}).
The four sets of points are, moving from left to right, linear
(\ref{eq:linearfit}), quadratic (\ref{eq:quadfit}), z (\ref{eq:zfit})
and pole (\ref{eq:monopole}) fits, respectively.
We observe that the systematic error due to the $q^2$ dependence is
comparable to that from the chiral extrapolation, and both are small
when compared to the statistical error.
Hence, one might hope that these systematic errors can be reduced
along with the statistical error by improving the statistics.

Finally, since we only have results at one lattice spacing, we are
unable to extrapolate to the continuum limit.
However, lattice artefacts are formally of $ O(a^2 \Lambda^2_{QCD})$;
assuming $\Lambda^2_{QCD} \sim 300$~MeV we, {\it tentatively}, estimate
these at $\approx 4\%$.
Hence our preliminary result is
\be
\Delta f = -0.0161(46)(15)(16)(7) \Rightarrow f_+^{K\pi}(0)=0.9609(51)\ ,
\ee
where the first error is statistical, and the second, third and fourth
are estimates of the systematic errors due to the chiral
extrapolation, $q^2$ dependence and lattice aritefacts, respectively.
This result agrees very well with the old value of Leutwyler \& Roos
\cite{Leutwyler:1984je} and earlier lattice calculations
\cite{Becirevic:2004ya,Okamoto:2004df,Tsutsui:2005cj,Dawson:2005zv,Dawson:2006qc}.

While this result for $f_+(0)$ is smaller than our earlier preliminary
analysis (0.9680(16)) \cite{Antonio:2006ev} and has a larger error, we
believe that through a more rigorous investigation of the systematics
involved in extracting this result, we are now much closer to having a
finalised result which we plan to publish soon.

Using $|V_{us}f_+(0)|=0.2169(9)$ from the experimental decay amplitude
\cite{blucher06}:
\be
|V_{us}| = 0.2257(9)_\text{exp}(12)_{f_+(0)}
\ee
we find
$
|V_{ud}|^2 + |V_{us}|^2 + |V_{ub}|^2 = 1-\delta,\quad \delta=0.00076(62)
$,
compared with the PDG(2006) result,
$\delta=0.0008(10)$.

\vspace*{-5mm}
\section{Summary and future work}
\vspace*{-3mm}
We have presented a preliminary result for $\Delta f= f_+(0) - (1 +
f_2)$ using $N_f=2+1$ dynamical domain wall fermions with three
choices for the light quark masses.
Our result $\Delta f = -0.0161(46)(15)(16)(7)$ agrees very well with
the Leutwyler \& Roos result \cite{Leutwyler:1984je} and confirms the
trend of other lattice results
\cite{Becirevic:2004ya,Okamoto:2004df,Tsutsui:2005cj,Dawson:2005zv,Dawson:2006qc}
which prefer a negative value for $\Delta f$.
We performed our simulations with matched parameters on two volumes
and we observe no obvious finite size effects.

We are in the process of improving this result by looking at ways of
decreasing the error on the point at $am_{ud}=0.01$.  We will also
soon have a result at a lighter quark mass $(am_{ud}=0.005)$ which
will assist in improving the chiral extrapolation.
Additionally, this result has been obtained at a single value of the
lattice spacing, so future simulations will need to be performed at
least at one more lattice spacing to investigate scaling behaviour.
\vspace*{-1mm}
\begin{acknowledgments}
  \vspace*{-3mm} 
  We thank D.\,Chen, N.\,Christ, S.\,Cohen, C.\,Cristian, Z.\,Dong,
  A.\,Gara, A.\,Jackson, C.\,Jung, R.\,Kenway, C.\,Kim, L.\,Levkova,
  X.\,Liao, G.\,Liu, R.\,Mawhinney, S.\,Ohta,
  K.\,Petrov, T.\,Wettig and A.\,Yamaguchi for developing
  with us the QCDOC machine and its software.  This development and
  the resulting computer equipment used in this calculation were
  funded by the U.S.\ DOE grant DE-FG02-92ER40699, PPARC JIF grant
  PPA/J/S/1998/00756 and by RIKEN.  This work was supported by DOE
  grants DE-FG02-92ER40699 and DE-AC02-98CH10886 and PPARC grants
  PPA/G/O/2002/00465, PP/D000238/1, PP/C504386/1, PPA/G/S/2002/00467
  and PP/D000211/1.
We thank BNL, EPCC, RIKEN, and the U.S.\ DOE for
supporting the computing facilities essential for the completion of
this work.
\end{acknowledgments}

\end{document}